%% file: neuro.tex
\pdfoutput=1

\documentclass[10pt,letter,twocolumn]{article}
\usepackage{f1000_styles}

\usepackage[super,comma,sort]{natbib}
\usepackage[colorlinks=true,urlcolor=blue]{hyperref}

\usepackage{amsmath}
\input neuro-def

\def\citeyear{\citep}
\def\autocite{\citep}
\def\textcite{\citet}

\begin{document}

\title{Puzzles in modern biology.~IV.~Neurodegeneration, localized origin and widespread decay}
\author[1]{Steven A.~Frank}
\affil[1]{Department of Ecology and Evolutionary Biology, University of California, Irvine, CA 92697--2525 USA, safrank@uci.edu}

\maketitle
\thispagestyle{fancy}

\parskip=4pt

\begin{abstract}

The motor neuron disease amyotrophic lateral sclerosis (ALS) typically begins with localized muscle weakness. Progressive, widespread paralysis often follows over a few years. Does the disease begin with local changes in a small piece of neural tissue and then spread? Or does neural decay happen independently across diverse spatial locations? The distinction matters, because local initiation may arise by local changes in a tissue microenvironment, by somatic mutation, or by various epigenetic or regulatory fluctuations in a few cells. A local trigger must be coupled with a mechanism for spread. By contrast, independent decay across spatial locations cannot begin by a local change, but must depend on some global predisposition or spatially distributed change that leads to approximately synchronous decay. This article outlines the conceptual frame by which one contrasts local triggers and spread versus parallel spatially distributed decay. Various neurodegenerative diseases differ in their mechanistic details, but all can usefully be understood as falling along a continuum of interacting local and global processes. Cancer provides an example of disease progression by local triggers and spatial spread, setting a conceptual basis for clarifying puzzles in neurodegeneration. Heart disease also has crucial interactions between global processes, such as circulating lipid levels, and local processes in the development of atherosclerotic plaques. The distinction between local and global processes helps to understand these various age-related diseases.

\end{abstract}


\vskip0.5in
\noterule
Preprint of published version: Frank, S. A. 2016. Puzzles in modern biology.~IV.~Neurodegeneration, localized origin and widespread decay. F1000Research 5:2537, \href{http://dx.doi.org/10.12688/f1000research.9790.1}{doi:10.12688/f1000research.9790.1}. Published under a Creative Commons \href{https://creativecommons.org/licenses/by/4.0/}{CC BY 4.0} license.
\smallskip
\noterule

\clearpage

\subsection*{Introduction}

Initial symptoms of motor neuron disease present as localized muscle weakness. Motor loss often progresses to widespread paralysis over a few years\autocite{baumer14advances}. 

The onset of this disease poses a puzzle. Does the disease arise in a localized focus of neural tissue and then spread from that focal lesion? Or does the decay arise independently in diverse spatial locations?

Suppose that disease begins from a localized origin\autocite{armon05acquired,armon16accrued,frank10somatic,prusiner12a-unifying,jucker13self-propagation}. Then onset may start by local changes in a tissue microenvironment, by somatic mutation, or by various epigenetic or regulatory fluctuations in a few cells. Those local processes may transform a small piece of tissue into a focal lesion that can spread disease to other cells. The widespread decay that ultimately follows happens by local transformation and then spread.

By contrast, suppose that widespread decay originates independently in each small site across the broad spatial domain of diseased tissue. Then localized genetic, epigenetic and regulatory changes in a single site cannot be the origin of the disease. Instead, spatially separated positions must progress independently. 

\subsection*{Clues from sporadic versus inherited disease}

Consider the pattern of onset and spread in the most common motor neuron disease, amyotrophic lateral sclerosis (ALS). 

The majority of cases occur sporadically\autocite{talbott16chapter}. Sporadic means that there is no direct evidence of predisposing inherited mutations. These apparently random cases typically occur after age 40, with incidence increasing up to age 75 and then declining at later ages\autocite{alonso09incidence}. 

Inherited mutations predispose individuals to ALS, causing familial occurrence\autocite{baumer14advances}. For example, individuals carrying an inherited mutation in \textit{SOD1} or \textit{C9orf72} often have greatly increased risk of disease. 

The age of onset in genetically predisposed cases typically occurs several years earlier than sporadic disease\autocite{cudkowicz97epidemiologya,ingre15risk}. Genetically predisposed individuals also have much higher incidence than those  without genetic predisposition.

The puzzle is whether disease begins with a local change that triggers global spread or with dispersed decay over a broad spatial range. The observed shift in age and incidence associated with inherited mutations provides clues.

Interpreting the clues from the age-incidence shift between familial and sporadic cases requires attention to two aspects. First, the puzzle concerns the dynamics of disease progression. To understand dynamics, we must consider the time-related aspects of the disease. Second, we must frame the clues in relation to the alternatives of localized versus dispersed origin.

\subsection*{Time from onset to full disease}

Individuals with certain inherited mutations have a high probability of developing ALS. However, the age at which symptoms first appear varies widely, even for carriers of the same mutation\autocite{cudkowicz97epidemiologya}. In sporadic cases, the age of first onset also varies widely.

Once initial symptoms arise, most individuals progress to final widespread paralysis within a few years. What could explain variable age for the first appearance of localized symptoms and the subsequent relatively rapid development of widely dispersed disease?

\subsection*{Localized versus dispersed origin}

I mentioned two possible solutions. First, disease may originate locally in a small piece of tissue and then spread from that origin. Second, degeneration may happen nearly simultaneously and independently across diverse spatial locations. 

The first solution of local origin and spread fits nicely with the observed pattern of variable age of onset and rapid subsequent progression. 

However, the second solution of parallel distributed decay could be true. For example, each individual might be prone to a particular timing of decay across the broad neural landscape. Approximate synchrony may arise because of the common genetic background or environmental exposures shared by all locations. 

For example, a global change in a widely circulating factor may initiate simultaneous decay across spatial locations. That global process shifts the locus of causality to the origin of the widely circulating trigger and to the susceptibility of the distributed sites across the neural landscape.

\subsection*{Trigger versus spread}

Inherited cases have an earlier age of onset than sporadic cases. That fact refines the alternative solutions of local versus dispersed origin\autocite{alonso09incidence,cudkowicz97epidemiologya,ingre15risk}.

In the local origin solution, a shared mutation across all locations may increase the rate at which the first localized origin arises. An origin may require several local changes before it can act as a trigger to initiate spatial spread. If all locations share a mutation that moves progression ahead, then the first trigger will happen at an earlier age.

Alternatively, the shared mutation across all locations may reduce the threshold for spread. A lower threshold may induce spread in response to a weaker local trigger.

\subsection*{Seed and soil}

A reduced threshold for spread suggests a variant of the dispersed origin solution. A reduced global threshold expresses distributed decay, but one that still requires an additional local origin trigger.

The interaction between local origin and dispersed decay echoes an old idea from cancer research about seed and soil\autocite{fidler03the-pathogenesis}. In that theory, the metastatic spread of cancer requires both a transformed cell that can act as a seed and a transformed tissue that can act as a soil in which the seed may grow.

\subsection*{Candidate mechanisms}

Alternative explanations focus attention on different mechanisms of disease.

Local triggers may arise from various processes: localized environmental insults, tissue microenvironment fluctuations such as infection or inflammation, local vascular changes, local hypoxia, and local changes in other kinds of environmental factors. Changes within one or few cells also initiate local changes: somatic mutation, epigenetic changes, fluctuations in regulatory state, phenotypic responses to altered environments, and so on. 

Spread may follow from intercellular transfer of RNA or cytoplasmic components, transmissible misfolding of proteins, diffusible signals, attraction of inflammatory responses, and so on. 

Dispersed origin may arise from wider environmental changes, including extrinsic insults, inflammation, broad vascular changes, and so on. 

Dispersed origin seems less likely to follow from localized somatic mutation, random epigenetic changes in cells, or random fluctuations in cellular regulatory states. This limitation and the absence of important mechanisms of spread provide the clearest distinction between local versus dispersed origin.

Much research focuses on these kinds of alternative mechanisms. However, mechanistic studies often do not explicitly frame analysis of cause in terms of the variety of potential mechanisms for local triggers and spread versus the variety of potential mechanisms for dispersed origin. My only purpose here is to clarify the relation between different mechanisms and the broader framework in which we must understand the puzzles of disease onset and progression.

In the study of mechanism, one must also distinguish rate of onset versus physiological function\autocite{frank16puzzles}. An inherited mutation may increase the rate at which disease-causing changes arise in physiological function, but the inherited mutation itself may have no direct physiological role in disease. 

For example, inherited defects in modulators of protein folding or in clearance of misfolded proteins may raise the rate at which misfolded proteins act as local triggers of global spread. Similarly, an inherited increase in somatic mutation may raise the rate at which local triggers arise.

Alternatively, an inherited mutation may directly initiate a disease-causing change in a physiological function. For example, a mutation in a protein coding gene may increase the tendency for misfolding of that particular protein. The increased tendency for misfolding may act as a local trigger or may lower the global threshold in response to external triggers. 

\subsection*{Neurodegenerative diseases}

I have used ALS to illustrate the puzzle of local versus dispersed origin of disease. Similar puzzles arise in Parkinson's disease, Alzheimer's disease and other neurodegenerative diseases. 

Within each disease, there will likely be different mechanisms of origin and timing of spread. Between diseases, there will also likely be different aspects of origin and spread. The similarities and differences help to understand broader aspects of disease.

\subsection*{Cancer}

At first glance, cancer and neurodegenerative disease seem very different. Cancer arises at a localized site. One thinks about the origin of cancer in terms of the local changes in a few cells and the surrounding tissue microenvironment. Global factors such as immune system status or hormone levels may play a role, but they do so to the extent that they influence local changes at the site of cancer origin. 

Progression of cancer depends on the factors that promote spread. The interactions between local triggers and global spread dominate all aspects of cancer research. The study of prevention, early detection, treatment, and basic understanding depends on the local-global interaction.

By contrast, most studies of neurodegeneration are vague about the origin and spread of disease. If a neurodegenerative disease does arise locally and then spread, then such a disease shares with cancer its general causal structure and dynamics.

Recently, several studies of neurodegeneration have focused on the spread of misfolded proteins in a prion-like manner\autocite{prusiner12a-unifying,jucker13self-propagation}. However, those studies remain vague about the variety of mechanisms that influence local triggers and about the broader conceptual framing of interactions between local and global processes.

Certainly, different neurogenerative syndromes vary in their causal structure, and various aspects of cancer and neurodegeneration differ in significant ways. It would be useful to understand explicitly the broad conceptual similarities and differences between the diseases. It would also be useful to understand the broader ways in which we can analyze the dynamics of interactions between local and global processes.

\subsection*{Heart disease}

Heart disease typically arises from an interaction of local and global processes. Initially, global factors such as lipid levels set the preconditions for localized plaque formation in the inner lining of artery walls. 

Although widespread conditions for plaque formation may occur, severe disease often requires a series of local changes at individual plaque sites\autocite{libby11progress}. For example, the early stages of local site progression typically include recruitment of leukocytes that mature into macrophages, which take up lipid. 

Changes in the local tissue microenvironment associate with proliferation of nearby muscle cells and tumor-like expansion and physiological transformation. An advanced plaque may rupture, attracting platelets and wound healing processes that make a clot. The clot may block local blood flow or break off to block flow at a distant site. 

Once again, a strong interaction between local and global processes drives disease progression. The particular timing of the local and global factors differs between heart disease, cancer and neurodegeneration. However, these age-related diseases share a common frame of interacting local and global processes that cause disease onset\autocite{doherty03calcification,shah15the-role}.

\subsection*{Conclusions}

Why does emphasis on interacting local and global processes matter? Consider the basic understanding of disease onset in neurodegeneration. 

If a local trigger starts the process, then a localized microenvironmental change or a local somatic mutation can be the event that initiates disease\autocite{armon05acquired,armon16accrued,frank10somatic}. By contrast, if a global change initiates disease, then we must look for a factor that can circulate or diffuse widely and that can alter conditions over dispersed spatial sites. 

With either initial local or global changes to start disease, progression typically depends on further interactions between subsequent local and global processes. For example, a high global level of certain lipids may be an important trigger of heart disease. Subsequent progression depends on local changes at plaque sites.

Much biological research hunts for the causes of disease. With better basic understanding of cause, one may improve prevention, detection and treatment. However, the notion of cause is always slippery and requires careful thought to frame properly. 

\subsection*{Competing interests}
No competing interests were disclosed.

\subsection*{Grant information}
National Science Foundation grant DEB--1251035 supports my research.

{\small\bibliographystyle{unsrtnat}
\bibliography{neuro}}

\end{document}

%% file: neuro-def.tex
\newcount\BoxNum \BoxNum 1\relax
\makeatletter
\newcommand*{\boxlabel}[1]{%
  \protected@write \@auxout {}{\string \newlabel {box:#1}{{\the\BoxNum}}{}}%
  \advance\BoxNum 1\relax}
\makeatother

\newcommand*{\boldrule}{\hrule height 1.2pt}
\newcommand*{\noterule}{\medskip\boldrule\medskip}	